\documentclass[pra,twocolumn,superscriptaddress,showpacs]{revtex4}
\usepackage{graphicx}  
\usepackage{dcolumn}   

\begin{document}

\def\be{\begin{equation}} 
\def\ee{\end{equation}}
\def\ba{\begin{eqnarray}} 
\def\ea{\end{eqnarray}} 

\title{Sympathetic
cooling and growth of a Bose-Einstein condensate}

\author{T. Papenbrock} 

\affiliation{Physics Division, Oak Ridge National Laboratory, Oak
Ridge, TN 37831, USA}

\author{A. N. Salgueiro} 

\affiliation{Max-Planck-Institut f\"ur Physik komplexer Systeme,
D-01187 Dresden, Germany}

\author{H. A. Weidenm\"uller} 

\affiliation{Max-Planck-Institut f\"ur Kernphysik, D-69029 Heidelberg,
Germany}

\date{\today}

\begin{abstract}

We study two sets of rate equations for sympathetic cooling of
harmonically trapped Bose gases. Calculations for mixtures of 
Na-Rb and Li-Cs show that both sets yield similar results for the 
cooling times. The equilibration rates are in fair agreement
with each other and differ considerably
from classical rates. The onset of Bose--Einstein condensation is
rather sudden and non--exponential in time, and the growth of the 
condensate differs for the two different mixtures we studied.

\end{abstract}

\maketitle

\section{Introduction}

The cooling of atoms in traps is an important tool in the study of
ultracold Bose or Fermi gases.  Usually, the last step in the cooling
process is evaporative cooling. This technique, however, can only be
used for sufficiently strongly interacting gases since thermodynamic
equilibrium of the cooled gas must be (nearly) attained during the
entire cooling process. There is a number of gases in which the
interaction is too weak for evaporative cooling to work. Moreover,
because of the exclusion principle the method fails at low temperature
for identical Fermions with parallel spins~\cite{fer}. In such cases,
one resorts to ``sympathetic cooling'': Another gas is cooled
evaporatively and acts as the cooling agent for the gas to be cooled.

The method of sympathetic cooling was proposed almost twenty years
ago\cite{win78,dru80} and has since found widespread application (see
Refs.~\cite{lar86,van85}). Exciting recent applications include the
sympathetic cooling of atoms of different species: ${}^{41}$K and
${}^{87}$Rb \cite{modugno}, ${}^7$Li and ${}^{144}$Cs \cite{mudrich},
of different isotopes of the same species: the bosonic species
${}^{85}$Rb cooled by ${}^{87}$Rb \cite{bloch} and ${}^6$Li Fermions
in a bath of ${}^7$Li Bosons \cite{truscott,schreck}, and the
production of dual Bose--Einstein condensates (BEC) with sympathetic
cooling\cite{myatt,delannoy}. Only a few theoretical descriptions of
sympathetic cooling exist. The simplest descriptions are purely
classical while more sophisticated employ the quantum Boltzmann
equation or master equations. Delannoy {\it et al.}~\cite{delannoy}
and Mosk {\it et al.}~\cite{Mosk} derived a classical formula for the
time dependence of the temperature and a classical formula for the
cooling rate.  Geist {\it et al.}~\cite{Geist} have formulated
sympathetic cooling of Fermi gases in terms of a quantum Boltzmann
equation. The tremendous simplification of the collision matrix
elements achieved in this way precludes, however, a quantitative
description of the cooling process.  Lewenstein {\it et
al.}~\cite{lew95} derived a master equation for sympathetic
cooling. Unfortunately, that master equation is too complex to be
practically useful. Recently, the present authors
simplified this master equation by means of decoherence and ergodicity
arguments~\cite{psw} and obtained two different sets of rate equations
that both describe sympathetic cooling of Bose or Fermi gases. These
rate equations are simple enough to permit practical calculations
while they contain detailed information about the quantum scattering
processes between the cooled gas and the bath.

The purpose of this article is twofold. First, we want to compare the
solutions of our rate equations with the much simpler expression for
the classical cooling rate obtained by Delannoy {\it et
al.}~\cite{delannoy} and Mosk {\it al.} \cite{Mosk}. 
Second, we use the rate equations to study the
growth of a Bose--Einstein condensate for a system of ${}^{23}$Na
cooled by ${}^{87}$Rb, and for ${}^7$Li in a bath of ${}^{133}$Cs.

This article is organized as follows. In Section~\ref{rates} we
present the rate equations that we use. The results, i.e. the
comparison between quantum and classical cooling rates and the study
of the growth of the condensate, are presented in
Sect.~\ref{result}. Finally we give our summary.

\section{Rate equations for sympathetic cooling}
\label{rates}

In this Section we present two different sets of rate equations that
are derived in ref.~\cite{psw}. Both sets of rate equations have been
derived from a master equation~\cite{lew95} for sympathetic cooling by
means of decoherence and ergodicity arguments. It is assumed that the
cooling agent has a considerably larger mass than the gas to be
cooled. The latter consists of $N_{A}$ (very weakly interacting)
bosons in a spherically harmonic trap of frequency
$\nu$. Single-particle trap orbitals have energies
$\varepsilon_{\vec{m}}=\hbar\nu(m_x+m_y+m_z)$ where
$\vec{m}=(m_x,m_y,m_z)$ denotes the set of Cartesian quantum numbers,
and the single-particle ground state energy is set to zero.

The first set of rate equations describes how the mean occupation number
$n_{\vec{m}_0}$ of orbital $\vec{m}_0$ changes due to interactions with 
the cooling agent:
\begin{eqnarray}
\label{fact}
\frac{\rm d}{{\rm d}t} n_{{\vec m}_0}(t) &=&  2 \sum_{{\vec m} (\neq
  {{\vec m}_0})} \Gamma^{{{\vec m}_0},{\vec m}}_{{\vec m},{{\vec m}_0}}
  n_{\vec m}(t) \left( n_{{\vec m}_0}(t) + 1 \right) \nonumber \\
&-& 2 \sum_{{\vec m} (\neq {{\vec m}_0})} \Gamma^{{\vec m},{{\vec
  m}_0}}_{{{\vec m}_0},{\vec m}} \left( n_{\vec m}(t) + 1 \right)
  n_{{\vec m}_0}(t) \ .
\end{eqnarray}
The form of these equations is intuitively obvious. Note that the
rate equations~(\ref{fact}) are of mean-field type and neglect
correlations between occupancies of different single-particle orbitals.
Particle number conservation implies that this
approximation may fail when the occupancy of the 
ground state $\vec{m}=0$ approaches
the total number of particles in the gas to be cooled. Particular
solutions to these mean-field type rate equations are discussed in
Refs.~\cite{MF}.

The rate coefficients $\Gamma^{{\vec m},{\vec n}}_{{\vec n},{\vec m}}$
are input to the rate equations~(\ref{fact}). For the computation of the
rate coefficients $\Gamma$ we assume that the bath
particles are Boltzmann-distributed.  A practical
recipe for their computation is given in Ref.\cite{psw}.  We express
the $\Gamma$'s in units of 
\be
\label{omega}
\omega=(32\pi^4)^{-1} \Lambda_B^3 n_B (a/
l_0)^2 [(M+m)^2 /(M m)]\nu.
\ee  
Here $n_B$ is the density of bath particles,
$\Lambda_B=(2\pi\hbar^2\beta_B/M)^{1/2}$ their thermal de--Broglie
wave length, $\beta_B$ the inverse temperature, $l_0$ the oscillator
length, and $M$ and $m$ are the mass of the cooling agent and the gas
to be cooled, respectively.  The interaction between bath and system
is described in terms of the $s$-wave scattering length $a$.

The second set of rate equations is based on a microcanonical approach
and describes how the probability $p_M$ of having the gas to be cooled
at total energy $M\hbar\nu$ changes with time: 
\ba
\label{micro}
\lefteqn{\frac{{\rm d} p_M}{{\rm d}t} = 2 \sum_{{\vec m} \neq {\vec
      n}} \Gamma^{{\vec m},{\vec n}}_{{\vec n},{\vec m}} }\nonumber\\
      && \times\biggl( p_{M+\alpha} \langle n_{\vec n} [n_{\vec m} +
      1] \rangle_{M+\alpha} - p_M \langle n_{\vec n} [n_{\vec m} + 1]
      \rangle_M \biggr) \ .  
\ea 
Here, $\alpha\hbar\nu=\varepsilon_{\vec n} -\varepsilon_{\vec m}$ is
the energy transfer, $n_{\vec m}$ is the occupancy of single-particle
orbital $\vec{m}$, and $\langle n_{\vec n}[n_{\vec m} + 1] \rangle_M$
denotes the mean value of $n_{\vec n} [n_{\vec m} + 1]$ taken over the
many-body states with fixed energy $M\hbar\nu$. They are thus
microcanonical averages of occupancies. It is straightforward to check
that $\sum_M {\rm d}p_M(t)/{\rm d}t = 0$. In Eq.~(\ref{micro}), the
$p_M$'s are the unknowns while the rate coefficients $\Gamma$ and the
quantities $\langle n_{\vec n}[n_{\vec m} + 1] \rangle_M$ are input
quantities defined in the framework of our model. The rate
coefficients are identical to those used in the mean-field type rate
equations~(\ref{fact}). To compute the quantities $\langle n_{\vec
n}[n_{\vec m} + 1] \rangle_M$ we neglect correlations by putting
$\langle n_{\vec n} [n_{\vec m} + 1]\rangle_{M} \approx \langle
n_{\vec n}\rangle_M [\langle n_{\vec m}\rangle_M + 1]$ and replace the
microcanonical average by the grand canonical average
\ba
\label{nocc}
\langle n_{\vec m}\rangle_M &=& n_{\vec m}(z,T)\nonumber\\
&=& {z(E)\exp{[-\varepsilon_{\vec m}/\kappa T(E)]}\over
1-z(E)\exp{[-\varepsilon_{\vec m}/\kappa T(E)]}}\bigg|_{E=M\hbar\nu} \ .
\ea
Expressions for the energy-dependent fugacity $z(E)$ and 
temperature $T(E)$ can be obtained numerically from the grand
canonical partition function of non-interacting bosons in a harmonic
trap.  For details, we refer the reader to Ref.\cite{psw}. Note that
the the microcanonical rate equations are expected to be valid also in
the regime of BEC.

\section{Results}
\label{result}

We study the cooling process for different temperatures of the bath
and for different combinations of bath and system. The system always
consists of $N_A=400$ Bosons and the harmonic trap is cut off beyond
the single-particle orbitals with energy $K\hbar\nu$, $K=21$.  
In the microcanonical approach, the
homogeneous system of linear rate equations~(\ref{micro}) can be put
into matrix form. For stability reasons, the real parts of the
eigenvalues of this matrix must be zero or negative. The equilibrium
solution is determined by the eigenvalue zero. The equilibration rate
is given by the modulus of the real part of the eigenvalue with
largest negative real part. The leading eigenvalues are computed using
the sparse matrix solver {\sc Arpack} \cite{Arpack}. Within our
numerical accuracy we found one zero eigenvalue and no eigenvalues
with positive real parts. The results for the equilibration rate
$\gamma_{\rm eq}$ are listed in Table~\ref{tab1}. The temperatures are
chosen such that $\kappa T=7\hbar\nu$ is close to the critical
temperature while $\kappa T=3\hbar\nu$ yields a BEC at the end of the
cooling process ($\kappa$ denotes Boltzmann's constant). 

We turn to the rate equations~(\ref{fact}). We took the Bosons of the
system $A$ to be initially equally distributed over the degenerate
single--particle orbitals with energy $K\hbar \nu$, all other orbitals
being empty. This initial configuration has the highest energy that can
be realized in the finite trap we are considering. While details of the
initial conditions are reflected in the short-time evolution they cannot
affect the long-time behavior and the equilibration rate. 
The rate equations~(\ref{fact}) need to be
modified in the regime of BEC; we therefore restrict the computation
to the case $\kappa T=7\hbar\nu$. We solve the rate equations
numerically and extract the equilibration rate $\tilde{\gamma}_{\rm
eq}$ from the solution at large times. The results are listed in
Table~\ref{tab1}. Note that the equilibration rates $\gamma_{\rm eq}$
and $\tilde{\gamma}_{\rm eq}$ depend only very weakly on the
composition of system and bath for $\kappa T=7 \hbar\nu$.  Comparing
the rate $\gamma_{\rm eq}$ obtained in the microcanonical approach
with the rate $\tilde{\gamma}_{\rm eq}$ from factorization, we note
that the former is about a factor $\approx 1.7$ larger than the
latter.  The time evolution however shows that the loss of energy at
short times is practically identical in both approaches. Most of the
energy is removed from system $A$ during the early part of the cooling
process.  Therefore, both approaches yield comparable predictions for
the cooling time.

\begin{table*}
\begin{ruledtabular}
\begin{tabular}{ldddd}
Bath - System & \kappa T/\hbar\nu & \gamma_{\rm eq} /  10^4\omega & 
\tilde{\gamma}_{\rm eq} /  10^4\omega & \bar{\gamma}_{\rm eq}/  10^4\omega \\\hline\hline
${}^{87}$Rb - ${}^{23}$Na & 7.0 & 2.7 & 1.6 & 5.84 \\\hline
${}^{87}$Rb - ${}^{23}$Na & 3.0 & 3.1 & -   & 1.07 \\\hline
${}^{133}$Cs - ${}^{7}$Li & 7.0 & 2.5 & 1.7 & 4.97 \\\hline
${}^{133}$Cs - ${}^{7}$Li & 3.0 & 1.66 & -  & 0.913 \\
\end{tabular}
\end{ruledtabular}
\caption\protect{\label{tab1}
Comparison of quantum--mechanical equilibration rates $\gamma_{\rm
  eq}$ (obtained from the solution of Eq.(\ref{micro})) and
  $\tilde{\gamma}_{\rm eq}$ (obtained from the solution of
  Eq.~(\ref{fact})) with the classical rate $\bar{\gamma}_{\rm eq}$
  for several systems and temperatures.  The critical temperature for
  onset of BEC is just below $\kappa T/\hbar\nu=7$.  The classical
  equilibration rates differ considerably from the quantum mechanical
  rates.}
\end{table*}

It is difficult to compare our full time dependence with that of the
classical theory \cite{delannoy,Mosk} since we assumed a finite trap
and, thus, different initial conditions. It is however meaningful
to compare equilibration rates since these do not depend on initial
conditions. In order to compare the
equilibration rates, we have adjusted the formulas of
Ref.\cite{delannoy,Mosk} to the case of a homogeneous bath. In
Table~\ref{tab1}, the classical rate $\bar{\gamma}_{\rm eq}$ (which
decreases quadratically with decreasing temperature) is seen to be
about a factor 2 larger than the quantum--mechanical rate $\gamma_{\rm
  eq}$) below the condensation temperature, and a factor 1.7 to 3
smaller in the regime of BEC. This casts some doubt on the possibility
to describe sympathetic cooling in terms of the much simpler classical
theory. 

\begin{figure}[b]
\includegraphics[width=6.5cm,angle=270]{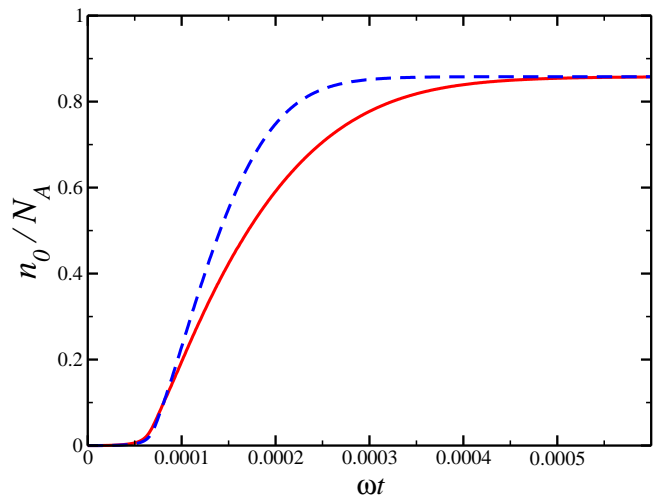}
\protect\caption{\label{fig1}Relative ground--state occupation as
function of time for Li in Cs bath (full line) and Na in Rb bath
(dashed line) at temperature $\kappa T=3\hbar\nu$.}
\end{figure}

It is particularly interesting to study the growth of a BEC. As
pointed out above, the rate equations~(\ref{fact}) cannot be used in
this regime.  We chose $1\kappa T=3\hbar\nu$ and followed the time
evolution of the total energy $E(t) = \hbar \nu \sum_M M p_M(t)$ by
solving the rate equations~(\ref{micro}) numerically. As initial
conditions we took $p_{K N_A} = 1$ and $p_{M} = 0$ for $M < KN_A$.
This is the configuration with highest energy that might be realized
in the open trap.
Using the grand canonical partition function, we obtain the
time-dependent temperature $T(E)$ for the system, and the
time-dependent ground state occupation number $n_0$ from
Eq.(\ref{nocc}). Fig.~\ref{fig1} shows the result for two different
systems. Note that the BEC grows faster in the Rb-Na system, which
also has the larger equilibration rate [in units of $\omega$, see
eq.~(\ref{omega})]. The reason for this behavior is not easy to
understand: Table~\ref{tab1} shows that equilibration rates for both
systems are quite similar at temperature $\kappa T/\hbar\nu = 7$ while
they differ considerably below the critical temperature. Note that a
naive (and classical) explanation would link the difference in rates
to the smaller momentum transfer during Li-Cs collisions and thereby
to the larger mass difference. Our quantum mechanical results show
that the situation is more complex. The onset of BEC is rather sudden
and non-exponential while the long time evolution is
exponential. Qualitatively similar observations have been made in
numerical simulations of evaporative cooling \cite{luiten,holland,gardiner}.
One does not obtain 100\% BEC since the final temperature is only
about half the condensation temperature and the system is not of
macroscopic size.

\section{Summary}

In summary, we have used two sets of rate equations to describe
sympathetic cooling of Bose gases by a much heavier gas. Practical
calculations in the systems ${}^{23}$Na-${}^{87}$Rb and
${}^7$Li-${}^{133}$Cs show that the cooling times from both rate
equations agree well with each other.  Our quantum--mechanical
equilibration rates are in fair agreement with each other but differ
considerably from the classical rates. This is unfortunate since the
classical approach is much simpler than the quantum--mechanical
description presented in this work. We studied the growth of the
Bose-Einstein condensate and found that the condensate in the system
${}^{23}$Na-${}^{87}$Rb grows faster than in ${}^7$Li-${}^{133}$Cs. At
the onset of Bose--Einstein condensation we observe a rather sudden
and non--exponential growth of the condensate.

\section*{Acknowledgments}
The authors are grateful to A. Mosk and M. Weidem\"uller for
suggesting this problem, and for many stimulating discussions. They
also thank S. J. Wang for valuable suggestions. T.P. thanks the 
Max--Planck--Institut f\"ur
Kernphysik, Heidelberg, for its hospitality during the initial stages
of this work. Oak Ridge National Laboratory is managed by UT-Battelle,
LLC for the U.S. Department of Energy under Contract
DE-AC05-00OR22725.


\begin{thebibliography}{99}

\bibitem{fer}
B. DeMarco {\it et al.},  
\prl {\bf 82} (1999) 4208.

\bibitem{win78}
D. J. Wineland, R. E. Drullinger, and F. L. Walls,
\prl {\bf 40} (1978) 1639.

\bibitem{dru80}
R. E. Drullinger, D. J. Wineland, and J. C. Berquist,
Appl. Phys. {\bf 22} (1980) 365.

\bibitem{lar86}
D. J. Larson {\it et al.}, 
\prl {\bf 57} (1986) 70.

\bibitem{van85}{\it Atomic Physics} 9, edited by R. S. van Dyck, Jr.,
  and E. N. Fortson, World Scientific, Singapore 1985.  

\bibitem{modugno} G. Modugno {\it et al.}, 
Science {\bf 294}, 1320 (2001).

\bibitem{mudrich} M. Mudrich {\it et al.}, 
arXiv: physics/0111213.

\bibitem{bloch} I. Bloch {\it et al.}, 
\pra {\bf 64}, 021402 (2001).

\bibitem{truscott}
A. G. Truscott {\it et al.}, 
Science {\bf 291} (2001) 2570.

\bibitem{schreck}
F. Schreck {\it et al.}, 
\pra {\bf 64} (2001) 011402(R).

\bibitem{myatt}
C. J. Myatt {\it et al.},  
\prl {\bf 78} (1997) 586.

\bibitem{delannoy}
G. Delannoy {\it et al.}, 
\pra {\bf 63} (2001) 051602(R).

\bibitem{Mosk}
A. Mosk {\it et al.}, 
Appl. Phys. {\bf B 73}, 791 (2001).

\bibitem{Geist}
W. Geist, L. You, and T. A. B. Kennedy,
\pra {\bf 59} (1999) 1500.

\bibitem{lew95}
M. Lewenstein, J. I. Cirac, and P. Zoller, 
\pra {\bf 51} (1995) 4617.

\bibitem{psw}
T. Papenbrock, A. N. Salgueiro, and H. A. Weidenm\"uller,
\pra {\bf 65}, 043601 (2002). 

\bibitem{MF}
 S. J. Wang, M. C. Nemes, A. N. Salgueiro, H. A. Weidenm\"uller,
\eprint cond-mat/0110293; \eprint cond-mat/0111413.

\bibitem{Arpack}
R. B. Lehoucq, D. C. Sorensen, and C. Yang,
{\it ARPACK User's Guide} 
(SIAM 1998) ISBN 0-89871-407-9.

\bibitem{luiten}
O. J. Luiten, M. W. Reynolds, J. T. M. Walraven,
\pra {\bf 53}, 381 (1996).

\bibitem{holland}
M. Holland, J. Williams, and J. Cooper,
\pra {\bf 55}, 3670 (1997).

\bibitem{gardiner} C. W. Gardiner {\it et al.}, \prl {\bf 79}, 1793 (1997);
C. W. Gardiner and P. Zoller, \pra {\bf 58}, 536 (1998);
C. W. Gardiner {\it et al.}, \prl {\bf 81}, 5266 (1998);
M. D. Lee and C. W. Gardiner, \pra {\bf 62}, 033606 (2000); M. K\"ohl 
{\it et al.}, \prl {\bf 88}, 080402 (2002).

\end{thebibliography}
\end{document}